\documentclass[a4paper,11pt]{article}
\pdfoutput=1 
\usepackage{jcappub} 
\usepackage[T1]{fontenc} 
\usepackage{amssymb,amsmath,amsfonts}
\usepackage{hyperref}
\usepackage{IEEEtrantools}
\usepackage{float}
\usepackage{physics}
\usepackage{braket}
\usepackage{simplewick}
\usepackage{graphicx}
\usepackage{xcolor}
\usepackage{tensor}
\usepackage[utf8]{inputenc}
\usepackage{caption,subfig}
\usepackage{bm,bbold}	
\usepackage{tikz}	
\usepackage{scalerel}
\usepackage{feynmf}
\usetikzlibrary{arrows,positioning,decorations.markings,decorations.pathmorphing,calc}
\usepackage[normalem]{ulem} 

\def\d{\mathrm{d}}

\newcommand{\para}[1]{\par\vspace{2mm}\noindent\emph{{#1}}.---}

\newcommand{\comment}[1]{}

\newcommand{\MPl}{M_{\rm P}}
\newcommand{\pd}{\partial}
\newcommand{\mr}[1]{\mathrm{#1}}
\def \be {\begin{equation}}
\def \ee {\end{equation}}
\title{Horndeski under the quantum loupe}
\author[a]{Lavinia Heisenberg,}
\author[b,c]{Johannes Noller,}
\author[a]{Jann Zosso}
\affiliation[a]{Institute for Theoretical Physics, 
ETH Zurich, Wolfgang-Pauli-Strasse 27, 8093, Zurich, Switzerland}
\affiliation[b]{Institute for Theoretical Studies, ETH Zurich, Clausiusstrasse 47, 8092 Zurich, Switzerland}
\affiliation[c]{DAMTP, University of Cambridge, Wilberforce Road, Cambridge CB3 0WA, U.K.}
\emailAdd{lavinia.heisenberg@phys.ethz.ch}
\emailAdd{johannes.noller@eth-its.ethz.ch }
\emailAdd{jzosso@phys.ethz.ch}
\abstract{With recent constraints on the propagation speed of gravitational waves, the class of scalar-tensor theories has significantly been reduced. We consider one of the surviving models still relevant for cosmology and investigate its radiative stability. The model contains operators with explicit breaking of the Galileon symmetry and we study whether they harm the re-organization of the effective field theory. Within the regime of validity we establish a non-renormalization theorem and show explicitly that the quantum corrections, to one-loop, do not detune the classical Lagrangian generating suppressed counterterms. This is striking since the non-renormalization theorem is established in the presence of a genuine Galileon symmetry breaking term.
}

\begin{document}
	\allowdisplaybreaks[1]
	\maketitle
	\flushbottom
	%

\section{Introduction}
\label{Sec:Introduction}
	
After the reassuring discovery of the Higgs boson we know that scalar fields exist in Nature, which was a tremendous step forward in particle physics. The question whether they could play an equally important role in cosmology is under intense investigation. The inflationary paradigm of the standard model of early universe cosmology and its successful realization enjoy the hypothetical presence of a scalar field. On an equal footing scalar fields could also play an important role for the late time universe in form of a dark energy candidate. Similarly, the standard model of particle physics accommodates both fundamental abelian and non-abelian vector fields as gauge forces. They might have relevant cosmological implications as well. 

Hence, a common practice in cosmology to address the inflation, the dark energy, and even the dark matter problem is to introduce additional degrees of freedom. Prominent classes include scalar-tensor \cite{Horndeski:1974wa}, vector-tensor \cite{Heisenberg:2014rta,Jimenez:2016isa}, tensor-tensor \cite{deRham:2010kj} and scalar-vector-tensor theories \cite{Heisenberg:2018acv}. Interestingly, all these theories contain Galileon interactions for the helicity-0 mode in certain limits. The Galileon model \cite{Nicolis:2008in} represents a scalar model with derivative self-interactions and equations of motion at most second order. As such, it inherently exhibits a Vainshtein-type screening mechanism when coupled to matter while avoiding any ghost pathologies. Its classical Lagrangian is technically natural in the sense, that the classical tuning of the coefficients is radiatively stable \cite{Luty:2003vm,Nicolis:2004qq,Hinterbichler:2010xn,dePaulaNetto:2012hm,Rham2013,Heisenberg:2014raa}. Other quantum aspects of the Galileon model were investigated in \cite{Brouzakis2014,Brouzakis2014a,Kampf2014,Pirtskhalava2015,Heisenberg:2019udf,Heisenberg:2019wjv}. Their unique quantum behaviour is not just a reflection of the Galileon symmetry $\pi\to\pi+c+b_\mu x^\mu$ but also their specific antisymmetric structure hidden behind the Levi-Civita tensors.

A new astonishing development in observational cosmology and astrophysics is the discovery of Gravitational Waves (GWs) as reported in 2016 by the LIGO team \cite{Abbott:2016blz}. A wealth of new exciting phenomena became accessible, among others, the interplay between GWs and the electromagnetic signal as observed for the GW170817 event \cite{TheLIGOScientific:2017qsa}. This new multimessenger data has enabled astrophysicists and cosmologists to put an incredibly tight constraint $\Delta c=10^{-15}$ on the propagation speed of GWs. Reluctance for dark energy models featuring an anomalous propagation speed was one of the immediate consequences \cite{Brax:2015dma,Lombriser:2016yzn,Ezquiaga:2017ekz,Creminelli:2017sry,Sakstein:2017xjx,Baker:2017hug} (see \cite{Heisenberg:2018vsk,Ezquiaga:2018btd,Kase:2018aps} for recent reviews). In this work we are interested in the Horndeski models, that have a luminal GW speed of propagation and hence trivially satisfy the GW170817 bound. We will call this subset the ``Horndeski survivals''\footnote{In particular we set aside other ways to satisfy the bound which involve introducing a frequency-dependent speed of GWs (in this context, especially see \cite{deRham:2018red}).} and will be curious about their validity as an effective field theory (EFT) under quantum corrections. Even if Horndeski models contain terms, that explicitly break the Galileon symmetry, their non-trivial mixing with Galileon invariant interactions on the vertices might soften the quantum behavior of the symmetry breaking terms. We will investigate these open questions using different specialized methods.

By finding the specific expansion parameters of the EFT based on power-counting arguments we analyze its structure and stability on all relevant scales. These results are then substantiated by the explicit calculation of various counterterm structures at one-loop on a flat background, cross-checked through the usage of three independent methods. With the minimal subtraction (MS) scheme (or its slightly modified version $\overline{\mathrm{MS}}$) in mind, we focus on the ultraviolet divergent (UV) part of the radiative corrections in dimensional regularization and directly compute all contributions up to the 4-point function and comment on higher order results.

The combination of the background field method with gauge preserving heat-kernel techniques provides powerful ways of computing the logarithmically divergent part of the effective action \cite{Schwinger1961,DeWitt1964,Atiyah1973,Abbott1982a,Barvinsky1985}. We will first perturbatively expand the effective action in traces expressed in a universal functional form, whose values are readily calculable by means of the generalized Schwinger-DeWitt formalism. Secondly, interpreting the background scalar field contribution to the second-order fluctuation term as an effective inverse metric enables one to define geometrical objects which bring the fluctuation operator into the form of a minimal second-order operator, necessary for the utilization of the original Schwinger-DeWitt technique. In this way, the divergent part of the one-loop effective action of all n-point functions are resummed in a single expression, from which individual contributions can be directly retrieved by expanding the curvature invariants in terms of the effective metric. Finally, the obtained results are again confirmed through standard Feynman momentum space calculations independent of the effective action. This provides us with strong confidence about the correctness of our results.

 These different methods were already applied in the context of the scalar Galileon model. The expansion of the one-loop effective action in terms of universal functional traces was used in \cite{dePaulaNetto:2012hm}, in order to calculate the correction to the two point function, as well confirmed by Feynman diagrammatic methods. The divergent part of the on-shell one-loop $4$-point function was investigated in detail in \cite{Kampf2014} and further generalized in \cite{Heisenberg:2019udf} beyond the on-shell limit up to one-loop $5$-point correlation functions. And in \cite{Heisenberg:2019wjv}, these results were extended to arbitrary $n$-point functions by the geometrical formulation described above.

The paper is organized as follows: In \S\ref{sec_HornSurv} we review what we call the ``Horndeski survivals'' in light of the GW170817 constraint, i.e. dark energy models that give rise to a luminal speed of propagation for GWs, and focus on one particular theory in this class. We first establish a non-renormalization theorem for our theory at hand using power counting arguments in \S\ref{Non-renormalization}. This is remarkable as the non-renormalization theorem is established in the presence of a Galileon symmetry breaking term. The one-loop quantum corrections are then computed on a flat background using first a perturbative generalized Schwinger-DeWitt technique up to the 4-point quantum corrections \S\ref{Schw} and then generalizing these by a geometrical resummation of all n-point function contributions in terms of an effective metric \S\ref{Geom}. The results are then tested against the computation of individual Feynman diagrams in \S\ref{sec_FeynDiag}. The obtained one-loop counterterms coincide with the power-counting arguments using dimensional analysis. The specific Horndeski survival is a viable EFT and does not receive quantum corrections.


\section{Surviving Horndeski Model}\label{sec_HornSurv}
Horndeski theories represent the most general scalar-tensor theories with three propagating degrees of freedom (two belonging to the spin-2 sector and one to the scalar field) with equations of motion at most second order.
The action is given by \cite{Horndeski:1974wa}
\begin{eqnarray}
\mathcal{S}=\int d^4x\sqrt{-g}\left(\sum_{i=2}^5\mathcal{L}_i+\mathcal{L}_{\rm matter}\right)
\end{eqnarray}
where the individual Horndeski Lagrangians read
\begin{eqnarray}\label{HorndeskiAct}
\mathcal{L}_2&=&G_2(\pi,X)\nonumber\\
\mathcal{L}_3&=&-G_3(\pi,X)[\Pi]\nonumber\\
\mathcal{L}_4&=&G_4(\pi,X)R+G_{4,X}\left([\Pi]^2-[\Pi^2]\right)\nonumber\\
\mathcal{L}_5&=&G_5(\pi,X)G_{\mu\nu}\Pi^{\mu\nu}-\frac16G_{5,X}\left([\Pi]^3-3[\Pi][\Pi^2]+2[\Pi^3]\right)\,.
\end{eqnarray}
Some words regarding the notation. The scalar kinetic term is denoted by $X=-\frac12(\partial\pi)^2$. The four arbitrary functions $G_2$, $G_3$, $G_4$ and $G_5$ depend on the scalar field $\pi$ and its kinetic term and their partial derivatives are represented by $G_{i,X}=\partial G_i/\partial X$ and $G_{i,\pi}=\partial G_i/\partial \pi$. Traces are symbolically expressed as $[\cdot]$. Special care is needed for the quartic and quintic interactions. The presence of non-minimal couplings to gravity via the Ricci scalar and the Einstein tensor is required in order for them to satisfy the above mentioned property of second order equations of motion. However, precisely these quartic and quintic interactions generate an anomalous propagation speed for GWs and their presence is tightly restricted.

For GWs propagating on cosmological backgrounds, consider small perturbations on top of a Friedmann-Lemaitre-Robertson-Walker (FLRW) metric $\bar{g}_{\mu\nu}=(-N(t)^2,a(t)^2\delta_{ij})$ and the background field configuration $\pi(t)$. The tensor perturbations take on a generic form dictated by the background symmetries
\begin{equation}
\mathcal{S}^{(2)}_T=\sum_{\lambda}\int d^4x a^3 q_T(\dot{h}_\lambda^2-\frac{c_T^2}{a^2}(\partial h_\lambda)^2)
\end{equation}
where $q_T$ stands for the modified gravitational coupling (frequently also referred to as an ``effective Planck mass'') and $c_T$ for the propagation speed \cite{Heisenberg:2018vsk,Kase:2018aps}
\begin{eqnarray}
q_T&=&\frac14\left( 2(G_4-2XG_{4,X})-2X(G_{5,X}\dot{\pi}H-G_{5,\pi}) \right) \nonumber\\
c_T^2&=&\frac{2G_4-2XG_{5,\pi}-2XG_{5,X}\ddot\pi}{4q_T}\,.
\end{eqnarray}
As apparent, the propagation speed of GWs within Horndeski theories depends on the background dynamics $H$, $\dot{\pi}$  and the background functions $G_4$, $G_5$. 
 For Horndeski theories, in the absence of an unnatural background tuning, the requirement of luminal propagation of GWs then translates into the condition
\begin{equation}\label{choices}
G_{4,X}=0 \qquad \text{and} \qquad G_5=0\,.
\end{equation}
These restrictions assure that the tensor perturbations on top of a cosmological background propagate with the speed of light $c_T=c$.
The remaining Horndeski interactions reduce to
\begin{equation}
S=\int d^4x \sqrt{-g} \left\{ G_2(\pi,X)-G_3( \pi,X)[\Pi]+G_4( \pi)R\right\}\,.
\end{equation}
As introduced and discussed in \cite{Noller:2018eht}, a relevant and interesting model (also for the linear cosmological perturbations) is 
\begin{equation} \label{Gdefs}
G_2=X,\qquad G_3=\frac{c_3}{\Lambda^3}X \qquad \text{and}  \qquad G_4=\frac{M_P^2}{2}\left( 1+\frac{c_4 \pi^2}{M_P\tilde{\Lambda}}\right) \,.
\end{equation}
Transformed into the Einstein frame, this model can explicitly be written as
\begin{equation} \label{exampleTheory}
S=\int \mathrm{d}^{4}x \sqrt{-g} \left[ \frac12M_P^2R+X\left(1-\frac{c_3}{\Lambda^3} \Box \pi+\frac{c_4^2 \pi^2}{\tilde{\Lambda}^2}\right)\right]\,,
\end{equation}
up to leading order in $1/\MPl$ and where we have absorbed a numerical ${\cal O}(1)$ factor into $\tilde\Lambda$. In \cite{Noller:2018eht} this setup was motivated by the observation that backgrounds with $\pi \sim \tilde \Lambda$ (where $\Lambda \ll \MPl$) can lead to ${\cal O}(1)$ deviations for linear perturbations from their standard $\Lambda{}$CDM cosmology evolution, while \eqref{Gdefs} ensures that the effective Planck mass is (to leading order) still just $\MPl$ in this case. In what follows we will somewhat decouple from this motivation though and simply take \eqref{exampleTheory} as an interesting example theory in its own right. Crucially for us, the theory includes both a Galilean invariant and a symmetry breaking one, where the symmetry breaking scale is $\tilde\Lambda$.

An important remark is in order. Certainly, Galileon interactions on curved spacetimes explicitly loose their invariance under the Galileon symmetry. Moreover, the explicit dependence of the Horndeski $G_4$ function on the scalar field considered here is another source of symmetry breaking. However, it is already known that shift symmetric Horndeski theories only weakly break the Galilean invariance and allow for quasi de Sitter backgrounds to be constructed, which are insensitive to loop corrections \cite{Pirtskhalava:2015nla}. Here we essentially investigate, whether some shift (and Galilean) symmetry breaking theories can enjoy similar features. 
Note that after going to the Einstein frame, the Horndeski scalar field will have a direct coupling to the standard matter fields and hence, there will be also loop contributions from the mixing with the matter fields. However, in this work we ignore the matter sector and fully concentrate on the Horndeski interactions.


\section{Non-renormalization theorem: Power-counting}\label{Non-renormalization}
As we mentioned above, the two sources of symmetry breaking are the Horndeski functions depending on the field value and the promotion to curved spacetime. In order to understand the role of the operators which do break the Galileon symmetry explicitly, the graviton sector will not be relevant at a first step. We are therefore interested in computing one-loop quantum corrections for the survival model on a flat background. Hence, we will consider the following action on top of a flat Euclidean space $g_{\mu\nu}=\delta_{\mu\nu}$
\begin{equation}
S=\int \mathrm{d}^{4}x\left(\mathcal{L}_{2}+\mathcal{L}_{3}+\mathcal{L}_{4}\right),\label{Gact}
\end{equation}
where the individual Lagrangian pieces are
\begin{align}
\mathcal{L}_{2}={}&\tilde{c}_2 \pi\epsilon^{\mu\nu\rho\sigma}\tensor{\epsilon}{^{\alpha}_{\nu\rho\sigma}}\,\partial_{\mu}\partial_{\alpha} \pi,\label{L1}\\
\mathcal{L}_{3}={}&\frac{\tilde{c}_3}{\Lambda^3} \pi\epsilon^{\mu\nu\rho\sigma}\tensor{\epsilon}{^{\alpha\beta}_{\rho\sigma}}\,\partial_{\mu}\partial_{\alpha} \pi\,\partial_{\nu}\partial_{\beta}\pi,\\
\mathcal{L}_{4}={}&\frac{\tilde{c}_4}{\tilde{\Lambda}^2} \pi^3\epsilon^{\mu\nu\rho\sigma}\tensor{\epsilon}{^{\alpha}_{\nu\rho\sigma}}\,\partial_{\mu}\partial_{\alpha} \pi\,,
\end{align}
i.e. we consider the scalar part of \eqref{exampleTheory} re-cast in such a way, that some of its anti-symmetric structure is made manifest via the Levi-Civita tensors. This will be useful later on. 
The terms $\mathcal{L}_{2}$ and $\mathcal{L}_{3}$ are invariant under Galilean symmetry $ \pi\to \pi+b_{\mu}x^{\mu}+c$ (up to total derivatives), but the last Lagrangian $\mathcal{L}_{4}$ explicitly breaks this symmetry. Note that we will use both Greek and Latin letters to denote space-time indices in what follows.
As concerns loop induced counterterms there will be three distinctive contributions:

\begin{enumerate}
\item
\textbf{Pure Galileon $\mathcal{L}_{3}$ insertions:} 
Pure Galileon interactions enjoy a very particular EFT structure. 
Counterterms arising from purely Galileon vertices from $\mathcal{L}_{3}$ will enjoy the known non-renormalization theorem: the Galileon symmetry, together with the antisymmetric structure of the interactions protect $\mathcal{L}_{3}$ from quantum corrections \cite{Luty:2003vm,Nicolis:2004qq,Hinterbichler:2010xn,Rham2013}. More precisely, all terms generated through quantum loops have more derivatives per field than the classical cubic galileon interaction and are dimensionally bound to the following schematic form\footnote{Each $\mathcal{L}_{3}$ vertex comes with a factor $1/\Lambda^3$, which fixes the number of derivatives per external field for a given number of vertices. Since there are no other scales in the theory the prefactor is also fixed, which accounts for higher loop contributions as well.} 
\be\label{Fgal}
\frac{\partial^{2n+4}}{\Lambda^{2n}}\left(\frac{\partial^2\pi}{\Lambda^3}\right)^m \sim \left(\frac{\partial^{2}}{\Lambda^{2}}\right)^{3+n}(\partial \pi)^2\left(\frac{\partial^2\pi}{\Lambda^3}\right)^{m-2} \,, \quad n\geq0\,,\;m\geq2\,.
\ee
This follows from the fact that only the log divergent piece enters in the construction of counterterms. Therefore, there exists a regime in the EFT below the UV cutoff, where the a priori irrelevant galileon operator becomes important compared to the kinetic term $\pd^2\pi\sim\Lambda^3$, while quantum corrections are still under control $\pd^2 \ll \Lambda^2$.  This defines the two expansion parameters
\be
\alpha_{\text{cl}}=\frac{\partial^2\pi}{\Lambda^3}\,,\quad \text{and}\quad \alpha_{\text{q}}=\frac{\partial^2}{\Lambda^2}\,.
\ee

A concrete, cosmologically relevant, example is the case where $\partial \sim H_0$ and $\pi \sim \MPl$, so that $\partial^2 \pi \sim \MPl H_0^2 \equiv \Lambda_3$ while $\partial^2/\Lambda_3^2 \sim (H_0/\MPl)^{2/3} \ll 1$.
In that sense, the EFT structure is very similar to GR, as one can have regimes with significant classical non-linearities, while quantum corrections are still under control. In the present context, this property is especially attractive with a possible cosmological application in mind, as it endows the theory very naturally with a Vainshtein-type screening effect. Since the pure Galileon interactions with relevant backgrounds will be under control, this sector will not be relevant for us.
\item
\textbf{Pure non-Galileon $\mathcal{L}_{4}$ insertions:} 
The $\mathcal{L}_{4}$ term, however, does not directly fit into the above organization of the EFT and we would like to study its implications for radiative stability. Note that this new interaction is irrelevant as well and will only be of importance as soon as $\pi^2\sim\tilde{\Lambda}^2$.
There will be counterterms generated through purely symmetry breaking non-Galileon vertices with $\mathcal{L}_{4}$ insertions. Based on dimensional analysis, the generated operators will be of the form 
\be\label{Fnew}
\pd^4\left(\frac{\pi}{\tilde{\Lambda}}\right)^{2l}\sim \frac{\pd^2}{\tilde{\Lambda}^2}(\partial \pi)^2\left(\frac{\pi^2}{\tilde{\Lambda}^2}\right)^{(l-1)}\,,\;l\geq 1\;.
\ee
An immediate consequence is that these contributions will not renormalize any of the classical operators. Furthermore, on the relevant scale $\pi^2\sim\tilde{\Lambda}^2$, one should expect to find a parametrically large regime, for which the generated quantum interactions are suppressed by a factor of $\frac{\pd^2}{\tilde{\Lambda}^2}$ compared to $\mathcal{L}_{4}$, thus introducing two new expansion parameters
\be
\alpha_{\tilde{\text{cl}}}=\frac{\pi^2}{\tilde{\Lambda}^2}\,,\quad \text{and}\quad \alpha_{\tilde{\text{q}}}=\frac{\partial^2}{\tilde{\Lambda}^2}\,.
\ee
Even though the number of derivatives generated via loop contributions are lower than the ones generated from the Galileon interactions, the additional derivatives $\frac{\pd^2}{\tilde{\Lambda}^2}$ at each n-point function is sufficient for the counterterms to be suppressed as long as $\pd^2 \ll \tilde{\Lambda}^2$.
\item
\textbf{Mixing of $\mathcal{L}_{3}$ and $\mathcal{L}_{4}$ insertions:} Finally, those counterterms induced by mixed vertices with insertions of both Galileon and non-Galileon interactions $\mathcal{L}_{3}$ and $\mathcal{L}_{4}$ generate counterterms which at one-loop go like
\be\label{Fmix}
\pd^4\left(\frac{\pi^2}{\tilde{\Lambda}^2}\right)^i\left(\frac{\pd^2\pi}{\Lambda^3}\right)^j\sim \frac{\pd^2}{\tilde{\Lambda}^2}(\partial \pi)^2\left(\frac{\pi^2}{\tilde{\Lambda}^2}\right)^{(i-1)}\left(\frac{\pd^2\pi}{\Lambda^3}\right)^j\,,\quad i,j\geq 1
\ee
where importantly we have made use of the fact that each $\mathcal{L}_3$ vertex contributes factors with at least two derivatives per leg -- a remnant of the Galileon non-renormalisation theorem discussed above, that remains true even when other symmetry-breaking operators are present as considered here.
Again, these will not generate any operators of the same form as the classical initial interactions. On scales for which both of these classical higher order self-interactions become relevant, there again exists a regime in which the quantum contributions are suppressed by the same parameter $\alpha_{\tilde{\text{q}}}$.
\end{enumerate}
Thus, the above powercounting arguments, which will be consoditated by the explicit calculation of quantum corrections in the following, already let us conclude, that at one-loop, all classical Lagrangian terms are protected against quantum corrections. Moreover, as long as $\alpha_{\text{q},\tilde{\text{q}}}\ll 1$ and $\alpha_{\text{cl},\tilde{\text{cl}}}\sim\mathcal{O}(1)$ or smaller, the quantum corrections remain suppressed, such that the restriction to observationally consistent theories will not be spoiled when coupling the theory to gravity.
At higher loop order, we expect the analysis to go through in parallel to the pure Galileon case, as these couterterm structures will merely be suppressed by additional factors of $\alpha_{\text{q}}$ and $\alpha_{\tilde{\text{q}}}$. 

The potentially worrisome expansion is not the expansion in loops, but rather the expansion in external legs. We have not yet mentioned the regime $\alpha_{\text{cl},\tilde{\text{cl}}}\gg 1$. Based on the expansions (\ref{Fgal},\ref{Fnew},\ref{Fmix}) one would need to conclude, that the EFT breaks down as terms with higher derivatives and higher numbers of background fields are considered, regardless of whether $\alpha_{\text{q},\tilde{\text{q}}}\ll 1$ or not. However, at a closer look this issue is cured, as was also realized in pure Galileon theories \cite{Nicolis:2004qq} and its possible UV completion via massive gravity \cite{deRham:2013qqa}. Namely, splitting the scalar field into it's background and fluctuation contribution one finds that the tree level kinetic term of quantum fluctuations gets enhanced by large classical-non-linearities, hence, precisely in the regime $\alpha_{\text{cl},\tilde{\text{cl}}}\gg 1$. As long as the classical contributions do not lead to ghost instabilities as is the case by construction, the quantum fluctuations are rather further suppressed on such scales in contrast to what one could have expected. In other words, upon canonical normalization the local cutoff gets effectively shifted towards the UV.

In summary, we can conclude that none of the classical operators will be renormalized and the non-renormalization theorem of pure Galileon interactions can be extended to the present theory. Moreover, the above analysis suggests that the general EFT organisation remains healthy on all of the relevant scales below the true UV cutoff and that the initial classical choice is technically a natural one.

The aim of the remainder of this paper is to explicitly compute the divergent part of the one-loop quantum corrections for the flat space model. In the following we will begin by presenting an explicit Schwinger-DeWitt calculation of the divergent one loop effective action up to the fourth order in background fields. 
After that, a geometrical interpretation of the second order differential operator will allow us to resum the contributions of all $n$-point functions into a single expression, which provides a closed algorithm for the calculation of one-loop counterterms to any order. We will then compare these results directly with the Feynman diagrammatic momentum space method.

But first, let's for a second go back to the initial action \eqref{exampleTheory}
\begin{equation}
S=\int \mathrm{d}^{4}x \sqrt{-g} \left[ \frac12M_P^2R+X\left(1-\frac{c_3}{\Lambda^3} \Box \pi+\frac{c_4^2 \pi^2}{\tilde{\Lambda}^2}\right)\right]\,.
\end{equation}
and argue that a restriction to flat space calculations is indeed justified, as the overall picture does not change in the full theory.
When coupled to gravity, the flat space-theory considered above is supplemented with an additional scale, the Plank scale $M_P$. From an EFT point of view, this will introduce a new class of mixed scalar-graviton diagrams through
\be \label{schematic_hp}
\sqrt{-g}X\sim(1+\tfrac{1}{M_P}h)(\pd\pi)^2\quad\text{and}\quad\nabla^2\pi\sim\pd^2\pi+\tfrac{1}{M_P}\pd h\pd\pi
\ee
where we have split $g_{\mu\nu}=\eta_{\mu\nu}+\frac{1}{M_P}h_{\mu\nu}$. However, resorting to the perfect viability of GR as an EFT, these new interactions will not spoil the hierarchy between classical and quantum contributions established above and we shall impose $\Lambda\ll M_P$ and $\tilde{\Lambda}\ll M_P$ in order to have a non-negligible observational effect on scales relevant for us. In our particular model, each power of the graviton field in the vertices of the resulting EFT Lagrangian comes with a heavy suppression of $1/M_P$. It is also worth to mention that our established non-renormalization theorem is very sensitive to the specific form of the assumed initial Lagrangian. Had we included more Galileon symmetry breaking terms in the Horndeski functions $G_2$ and $G_3$, establishing a similar non-renormalization theorem might have been difficult. Including non-minimal graviton couplings might also generate non-trivial mixings between the different scales, specially through non-trivial $X$-dependent $G_4$ function, even though they inevitably would violate the luminal GW speed of propagation in this case.


\section{One-loop effective action \'a la Schwinger-DeWitt technique}
\label{Schw}
The bedrock of the computation of the one-loop effective action is the background field method. 
For this, the Galileon field is split into it's classical background and small quantum fluctuation
\begin{align}\label{fieldSplit}
\pi(x)= \bar{\pi}(x)+\delta \pi(x)\,.
\end{align}
The Euclidean one-loop effective action is then given by
\begin{align}\label{EffA}
\Gamma_{1}=\frac{1}{2}\text{Tr}\log F(\partial)\,,
\end{align}
where the general form of the scalar second order differential operator reads
\begin{align}
F(\partial^x)\delta(x,x')=\left.\frac{\delta^2 S_{\mathrm{G}}[ \pi]}{\delta \pi(x)\delta \pi(x')}\right|_{\pi=\bar{\pi}}\,.\label{SecOrdOp}
\end{align}
The one-loop counterterms induced by the action \eqref{Gact} up to a given order in fields and derivatives can be obtained by calculating the logarithmic divergent part of the one-loop effective action \eqref{EffA} in the background field approach using the generalized Schwinger-DeWitt technique \cite{Barvinsky1985}.  

\subsection{Fundamental operators and expansions}
This method starts by splitting the scalar second order differential operator \eqref{SecOrdOp} into it's principle part $\Delta$ and the subleading, background field dependent perturbation $Y=Y( \bar{\pi})$
\be
\label{split}
F(\partial)=\Delta+Y\, ,
\ee
with
\begin{align}
\Delta={}&-\delta^{a b}\pd_a\pd_b\,,\\
Y={}&\tilde{c}_3\,\frac{12}{\Lambda^3}\left( \Delta  \bar{\pi}\Delta-\pd^a\pd^b  \bar{\pi}\pd_a\pd_b\right)-\tilde{c}_4\,\frac{36}{\tilde{\Lambda}^2}\left( \bar{\pi}^2\Delta+ \bar{\pi}\Delta \bar{\pi}\right)\, , \label{Y}
\end{align}
where we have canonically normalized by setting $\tilde{c}_2=-\frac{1}{12}$. Note that \eqref{Y} contains the contributions proportional to $\tilde{c}_3$ and $\tilde{c}_4$.

The splitting \eqref{split} together with an expansion of the logarithm in \eqref{EffA} leads to
\be
\label{logExp}
\scaleto{\frac{1}{2}\Tr \ln F(\nabla)=\frac{1}{2}\Tr \ln\left[ \Delta\right]+\frac{1}{2}\Tr\left[Y\frac{1}{\Delta}\right]-\frac{1}{4}\Tr\left[Y\frac{1}{\Delta}Y\frac{1}{\Delta}\right]+\frac{1}{6}\Tr\left[Y\frac{1}{\Delta}Y\frac{1}{\Delta}Y\frac{1}{\Delta}\right]+\mathcal{O}(Y^4)\mathstrut}{24pt}\, ,
\ee
where $\frac{1}{\Delta}$ denotes the inverse of the principle operator.

The method now consists of transforming the expansion above into a sum of terms proportional to universal functional traces whose divergent part can readily be evaluated. In flat spacetime, the only non-vanishing universal functional traces in dimensional regularization with $d=4-2\epsilon$\footnote{Note that we have not carried around the various factors of $d$ arising when converting the Levi-Civita structure in the Lagrangian \eqref{L1} to contractions of the metric tensor, since the divergent part at one loop is blind to the extra $\epsilon$ terms. Moreover, the theory could have been defined from the start without explicit use of any Levi-Civita symbol.} have the form
\be
\label{UFT}
\Tr\;\mathcal{Y}^{\mu_{\scaleto{1\mathstrut}{4pt}}...\mu_{\scaleto{2n-4\mathstrut}{4pt}}}(\bar{\pi})\,\pd_{\mu_{\scaleto{1\mathstrut}{4pt}}}...\pd_{\mu_{\scaleto{2n-4\mathstrut}{4pt}}}\,\frac{1}{\Delta^n}\bigg\rvert_{\text{div}}=\,\frac{(-1)^n}{16 \bar{\pi}^2\,\epsilon}\,\int\mathrm{d}^{4}x\,\mathcal{Y}^{\mu_{\scaleto{1\mathstrut}{4pt}}...\mu_{\scaleto{2n-4\mathstrut}{4pt}}}(\bar{\pi})\,\frac{\delta^{(n-2)}_{\mu_{\scaleto{1\mathstrut}{4pt}} ... \mu_{\scaleto{2n-4\mathstrut}{4pt}}}}{2^{n-2}\,(n-1)!}\, ,
\ee
where $n\geq 2$ and  $\delta^{(n-2)}_{\mu_{\scaleto{1\mathstrut}{4pt}} ... \mu_{\scaleto{2n-4\mathstrut}{4pt}}}$ is the totally symmetrized product of $n-2$ metrics. Observe that the background field dependent piece $\mathcal{Y}(\bar{\pi})$ just goes along the ride, regardless of it's specific form. 

Any term appearing in the expansion \eqref{logExp} can be cast into the specific form appearing on the left hand side of \eqref{UFT} by commuting all the operators $\frac{1}{\Delta}$ to the right. Note that
\be \label{Com1}
\scaleto{\left[\frac{1}{\Delta}\,,Y\right]=-\,\frac{1}{\Delta}\,[\Delta\,,Y]\,\frac{1}{\Delta}\mathstrut}{24pt}\, ,
\ee
where each commutation increases the number $n$ of inverse operators $\frac{1}{\Delta}$ as well as the number of derivatives acting on the background operator Y
\be
 [\Delta\,,Y]=(\Delta Y)-2(\pd^\alpha Y)\pd_\alpha \, .
\ee
Given that one is only interested in counterterms up to a given order in the fields as well as a given order in derivatives applied to them, the procedure above is efficient in the sense that the log expansion \eqref{logExp} will be cut off by the maximum number of background fields one is interested in, while the iterative commutation of operators \eqref{Com1} will eventually hit the threshold of derivatives applied on the background fields, such that all traces indeed can take on a universal functional form \eqref{UFT}.


\subsection{Results up to 4-point function}

In our case we will compute the logarithmic divergent part of the one-loop effective action up to four background fields, that is the 4-point function contributions, acted on by a maximum of ten derivatives, which translates into a limitation to ten external momentas. 

First of all, note that from \eqref{UFT} it follows that the linear terms $\Tr\left[Y\frac{1}{\Delta}\right]$ with $n=1$ remain finite in dimensional regularization and can thus be disregarded. This directly implies that the 1-point tadpole contribution and the 2-point contribution proportional to $\tilde{c}_4$ do not contribute.

The next term in the log expansion \eqref{logExp} $\sim Y^2$ give rise to an already known, pure galilean contribution to the 2-point function\footnote{see eg. \cite{dePaulaNetto:2012hm,Heisenberg:2019udf,Heisenberg:2019wjv}} and new contributions to the 3- and 4-point functions proportional to $\tilde{c}_3 \tilde{c}_4$ and $\tilde{c}_4^2$ respectively:
\begin{align}
\Gamma_{1,3}^{\rm div} &\supset - \frac{54}{16 \pi^2\epsilon}\,\frac{\tilde{c}_3\tilde{c}_4}{\Lambda^3\tilde{\Lambda}^2}\,\int \mathrm{d}^4x \,  \bar{\pi}\,\Delta  \bar{\pi}\,\Delta^2  \bar{\pi}\, ,\label{resultc3c4} \\
\Gamma_{1,4}^{\rm div} &\supset - \frac{324}{16 \pi^2\epsilon}\,\frac{\tilde{c}_4^2}{\tilde{\Lambda}^4}\,\int \mathrm{d}^4x \,  \bar{\pi}^2\,(\Delta  \bar{\pi})^2\label{resultc4c4}\, .
\end{align}
The concise form of the above results can be obtained by performing several tuned integrations by parts and the equivalence to more basic results can conveniently be checked by going into momentum space which eliminates this freedom of representation (see for instance \S\ref{sec_FeynDiag}).

In the same spirit, the $\sim Y^3$ term in \eqref{logExp} will yield a known contribution $\sim \tilde{c}_3^3$ to the 3-point function and a novel mixed contribution $\sim \tilde{c}_3^2\tilde{c}_4$ to the 4-point function, while other contributions will depend on more than four background fields. The next order will then merely contribute to the 4-point function via a pure galileon contribution $G(\tilde{c}_3^4)$ which we are not interested in here. The final results up to the fourth order in background fields read
\begin{IEEEeqnarray}{rCl}\label{finalResults}
\Gamma_{1,2}^{\rm div} &=\frac{-1}{16 \pi^2\epsilon}\,\int \mathrm{d}^4x &\, \frac{9}{4}\,\frac{\tilde{c}_3^2}{\Lambda^6}\,  \bar{\pi}\,\Delta^4  \bar{\pi}\,,\nonumber\\
\Gamma_{1,3}^{\rm div} &=\frac{1}{16 \pi^2\epsilon}\,\int \mathrm{d}^4x \, &\left[\frac{\tilde{c}_3^3}{\Lambda^9}\left\{\tfrac{63}{4} \Delta{} \bar{\pi} (\Delta^2{} \bar{\pi})^2+ \tfrac{9}{2}(\Delta{} \bar{\pi})^2 \Delta^3{} \bar{\pi}  - \tfrac{9}{2}  \bar{\pi}\Delta^2{} \bar{\pi}\Delta^3{} \bar{\pi}\right.\right.\nonumber\\
&&\left.\left.- \tfrac{9}{4} \bar{\pi}  \Delta{} \bar{\pi} \Delta^4{} \bar{\pi} + \tfrac{27}{8}  \bar{\pi}^2 \Delta^5{} \bar{\pi} \right\}-54\,\frac{\tilde{c}_3\tilde{c}_4}{\Lambda^3\tilde{\Lambda}^2}  \bar{\pi}\,\Delta  \bar{\pi}\,\Delta^2  \bar{\pi}\right]\nonumber\, , \\
\Gamma_{1,4}^{\rm div}& = \frac{-1}{16 \pi^2\epsilon}\,\int \mathrm{d}^4x&\left[\frac{\tilde{c}_3^2\tilde{c}_4}{\Lambda^6\tilde{\Lambda}^2}\{\tfrac{657}{10} (\Delta{} \bar{\pi})^4 -  \tfrac{2727}{2} \bar{\pi}(\Delta{} \bar{\pi})^2 \Delta^2{} \bar{\pi}  + \tfrac{1134}{5}\bar{\pi}^2 (\Delta^2{} \bar{\pi})^2 -\tfrac{666}{5}\bar{\pi}^2 \Delta{} \bar{\pi} \Delta^3{} \bar{\pi}  \right.\nonumber\\
&&\left.
+\bar{\pi}^3 \tfrac{114}{5} \Delta^4{} \bar{\pi}  +  \tfrac{3969}{5} \bar{\pi}\Delta{} \bar{\pi}   \partial_{a}\Delta{} \bar{\pi}\partial^{a}\Delta{} \bar{\pi}+ 432 \Delta{} \bar{\pi}\Delta^2{} \bar{\pi}\partial_{a} \bar{\pi}\partial^{a} \bar{\pi}\right.\nonumber\\
&&\left.
 + \tfrac{2592}{5} \Delta{} \bar{\pi}\partial_{b}\partial_{a}\Delta{} \bar{\pi}\partial^{a} \bar{\pi}\partial^{b} \bar{\pi}+\tfrac{216}{5}(\bar{\pi} \Delta^2{} \bar{\pi} -  2 (\Delta{} \bar{\pi})^2) \,\partial_{b}\partial_{a} \bar{\pi}\partial^{b}\partial^{a} \bar{\pi}\right.\nonumber\\
&&\left.
+\tfrac{468}{5} \bar{\pi}\Delta{} \bar{\pi} \partial_{c}\partial_{b}\partial_{a} \bar{\pi} \partial^{c}\partial^{b}\partial^{a} \bar{\pi}+\tfrac{1152}{5} \Delta{} \bar{\pi}\partial^{b}\partial^{a} \bar{\pi} \partial_{c}\partial_{b} \bar{\pi} \partial^{c}\partial_{a} \bar{\pi}\}\right.\nonumber\\
&&\left.
+324\,\frac{\tilde{c}_4^2}{\tilde{\Lambda}^4}\,  \bar{\pi}^2\,(\Delta  \bar{\pi})^2\, + G(\tilde{c}_3^4)\right]\,.
\end{IEEEeqnarray}
\\

\noindent The contributions coming from purely Galileon interactions coincide with the known results in the literature (for instance with \cite{Heisenberg:2019wjv}). We see exactly that our dimensional analysis performed in \S\ref{Non-renormalization} is directly reflected in the individual counterterms generated at one loop. For instance, the three point function of the pure Galileon interactions proportional to $\tilde{c}_3^3$ generates an operator involving 10 derivatives compared to the classical $\mathcal{L}_{3}$ Lagrangian with 4 derivatives. This counterterm is suppressed as long as $ \alpha_{\text{q}}=\frac{\partial^2}{\Lambda^2}\ll1$ and the large number of derivatives generated is at the heart of the well-known non-renormalization theorem of the Galileon. Interestingly, we also see this non-renormalization property for the pure symmetry breaking and mixed contributions calculated above, as already anticipated by the dimensional analysis in section \ref{Non-renormalization}. Explicitly, the correction to the four point function originating from the symmetry breaking interaction proportional to $\tilde{c}_4^2$ yields a contribution with four derivatives applied on the four fields, while the classical $\mathcal{L}_{4}$ Lagrangian only involves two. Hence, the non-renormalization holds and the generated counterterms remain suppressed assuming $\alpha_{\tilde{\text{q}}}=\frac{\partial^2}{\tilde{\Lambda}^2}\ll1$ in this case.
The same is true for the mixed counterterms, i.e. proportional to combined powers of $\tilde{c}_3$ and $\tilde{c}_4$. The contribution in $\Gamma_{1,3}^{\rm div}$ proportional to $\tilde{c}_3\tilde{c}_4$ and the one proportional to $\tilde{c}_3^2\tilde{c}_4$ in $\Gamma_{1,4}^{\rm div}$ also give rise to counterterms involving two more derivatives as compared to the classical Lagrangian. This can be viewed as a remnant of the pure galileon non-renormalization theorem.

Summarizing, we conclude that our specific, cosmologically relevant Horndeski survival model shares a non-trivial non-renormalization theorem even in the presence of symmetry breaking operators.


\section{Closed algorithm: Geometrical resummation}\label{Geom}

We will now proceed and present a closed algorithm for the calculation of the divergent one-loop effective action to any order. On the one hand, this will give a non-trivial check of the above results and on the other it will allow us to have access to arbitrary higher order terms.
Similar to the previous section \ref{Schw} we split the Galileon field into it's background and perturbation part as in equation \ref{fieldSplit}. The one-loop effective action is again given by $\Gamma_{1}=\frac{1}{2}\text{Tr}\log F(\partial)$. 
This time we represent the scalar second order differential operator as
\begin{align}\label{DiffOpGeom}
F(\partial^x)\delta(x,x')=\left.\frac{\delta^2 S_{\mathrm{G}}[ \bar{\pi}]}{\delta \bar{\pi}(x)\delta \bar{\pi}(x')}\right|_{\pi=\bar{\pi}}=\left(-M^{\mu\nu}\partial_{\mu}^{x}\partial_{\nu}^{x}+\tensor{\Gamma}{^{\nu}}\partial_{\nu}^{x}+P\right)\delta(x,x')\,.
\end{align}
For the theory at hand \eqref{Gact} the  explicit contributions are
\begin{align}
&M^{\mu\nu}={}-\left(2\tilde{c}_2\,\varepsilon^{\mu\alpha\rho\sigma}{{\varepsilon}^{\nu}}_{\alpha\rho\sigma}+6\frac{\tilde{c}_3}{\Lambda^3}{\varepsilon}^{\mu\alpha\rho\sigma}{{\varepsilon}^{\nu\beta}}_{\rho\sigma}\,\partial_{\alpha}\partial_{\beta} \bar{\pi} +6\color{black} \bar{\pi}^2\frac{\tilde{c}_4}{\tilde{\Lambda}^2} \varepsilon^{\mu\alpha\rho\sigma}{{\varepsilon}^{\nu}}_{\alpha\rho\sigma}\right)\,,\label{expSOO} \\
&\tensor{\Gamma}{^{\nu}}={}0 \, ,\nonumber\\
&P={} 6 \bar{\pi}\frac{\tilde{c}_4}{\tilde{\Lambda}^2}\varepsilon^{\mu\alpha\rho\sigma}{{\varepsilon}^{\nu}}_{\alpha\rho\sigma}\partial_{\mu}\partial_{\nu} \bar{\pi}\, .\label{expSOOP}
\end{align}
Recall that we are working in Euclidean space, so $\varepsilon_{\mu\nu\rho\sigma}\varepsilon^{\mu\nu\rho\sigma} = d!$ (where $d=4$ for us) and setting $c_2 = -1/12$ canonically normalises the kinetic term.
Note that in the absence of the Galileon symmetry breaking interaction $\mathcal{L}_{4}$ only the symmetric tensor $M^{\mu\nu}$ would contribute, which has been discussed in detail in \cite{Heisenberg:2019wjv}.

The algorithm starts by identifying the symmetric tensor $M^{\mu\nu}$ as the inverse of an effective metric $M_{\mu\nu}$, such that 
\be\label{defeffM}
M_{\mu\rho}M^{\rho\nu}=\delta^{\mu}_{\nu}\,,
\ee
assuming that the effective metric is non-degenerate $M\equiv \det [M_{\mu\nu}]\neq 0$. The effective metric $M_{\mu\nu}$ then allows the definition of a corresponding metric compatible covariant derivative $\nabla^{M}_{\mu}$ with associated connection $\tensor{\Gamma}{^{\rho}_{\mu\nu}}(M)=\frac{M^{\rho\sigma}}{2}\left(\partial_{\mu}M_{\sigma\nu}+\partial_{\nu}M_{\sigma\mu}-\partial_{\sigma}M_{\mu\nu}\right)$,
such that $\nabla^{M}_{\rho}M_{\mu\nu}=0$. This provides us with an effective Laplacian $\Delta_{M}:=-M^{\mu\nu}\nabla^{M}_{\mu}\nabla^{M}_{\nu}$, with which we can reformulate the first term in \eqref{DiffOpGeom}: $-M^{\mu\nu}\partial_{\mu}\partial_{\nu}=\Delta_{M}-M^{\mu\nu}\tensor{\Gamma}{^{\rho}_{\mu\nu}}(M)\nabla^{M}_{\rho}$.

Thus, the operator \eqref{DiffOpGeom} can be rewritten in terms of quantities defined through the effective metric as
\begin{align}\label{Op2}
F(\nabla^{M})=\Delta_{M}-2L^{\rho}\nabla_{\rho}^{M}+P\,,
\end{align} 
where $L^{\rho}$ is defined to be
\begin{align}
L^{\rho}\equiv\frac{1}{2}M^{\mu\nu}\tensor{\Gamma}{^{\rho}_{\mu\nu}}(M)\,.
\end{align} 

Finally, by redefining the covariant derivative $\mathcal{D}_{\mu}:=\nabla^{M}_{\mu}+M_{\mu\nu}L^{\nu}$, the second-order fluctuation  operator \eqref{SecOrdOp} can be brought into a minimal second order form
\begin{align}\label{Opmin}
F(\mathcal{D})=-\mathcal{D}_{\mu}\mathcal{D}^{\mu}+U\,,
\end{align}
where all the linear terms have been absorbed by the potential part
\begin{align}\label{pot}
U\equiv\nabla_{\nu}^{M}L^{\nu}+L_{\nu}L^{\nu}+P\,.
\end{align}

Using heat-kernel techniques, the one-loop divergences of the effective action \eqref{EffA} can then be expressed in a closed form in terms of geometrical curvature invariants of the effective metric $M_{\mu\nu}$ and the potential $U$ \cite{Heisenberg:2019wjv}
\begin{IEEEeqnarray}{rCl}\label{OneLoopGgen}
\Gamma_{1}^{\mathrm{div}}&=-\frac{\chi(\mathcal{M})}{180\varepsilon}-\frac{1}{32 \bar{\pi}^2\varepsilon}\int_{\mathcal{M}}\mathrm{d}^4x\,\sqrt{M}\,&\left\{\frac{1}{60}M^{\mu\rho}M^{\nu\sigma}R_{\mu\nu}(M)R_{\rho\sigma}(M)\right.\nonumber\\
&&\left.\;+\frac{1}{120}R^2(M)-\frac{1}{6}R(M)U+\frac{1}{2}U^2\right\}\,,
\end{IEEEeqnarray}
where $\chi(\mathcal{M})=\frac{1}{32 \pi^2}\int_{\mathcal{M}}\mathrm{d}^4x\,\sqrt{M}\,\mathcal{G}(M)$ is the Euler characteristic of $\mathcal{M}$ in $d=4$ dimensions in terms of the Gauss-Bonnet $\mathcal{G}(M)=R_{\mu\nu\rho\sigma}(M)R^{\mu\nu\rho\sigma}(M)-4R_{\mu\nu}(M)R^{\mu\nu}(M)+R^2(M)$. However, since the effective metric is symmetric and metric compatible, the Gauss-Bonnet term can be discarded in four dimensions (we have explicitly checked that all resulting $\bar \pi$ interactions are indeed total derivatives, as expected) and we are thus left with
\begin{align}\label{OneLoopGour}
\Gamma_{1}^{\mathrm{div}}=-\frac{1}{32 \pi^2\varepsilon}\int_{\mathcal{M}}\mathrm{d}^4x\,\sqrt{M}\left\{\frac{R_{\mu\nu}R^{\mu\nu}}{60}
+\frac{R^2}{120}-\frac{RU}{6}+\frac{U^2}{2}\right\}\,.
\end{align}

In order to extract one-loop counterterms from the full resummed result \eqref{OneLoopGour} one simply plugs in the effective metric and its inverse and expands up to the desired order of background fields $ \bar{\pi}$. The explicit expressions of the effective metric and its determinant up to $4^{\text{th}}$ order in the field $ \bar{\pi}$ can be found in the Appendix \ref{explicityGeom}. In this way, all geometrical objects defined above can be expanded in the number of background fields such that the method provides a closed algorithm for the calculation of all the one-loop counterterms of the theory.
Doing so to the required orders in $\bar\pi$, we indeed precisely recover all expressions in \eqref{finalResults}.

However, the same can be obtained by resorting to metric perturbation tools without ever needing to perturbatively invert the effective inverse metric $M^{\mu\nu}$. We refer the reader to section 4 in \cite{Heisenberg:2019wjv} for more details. First of all, we expand the effective metric employed in the geometrized formulation to a desired order $n$
\be\label{Mexp}
M_{\mu\nu}=\delta_{\mu\nu}+\sum_{l=1}^n h^{\scaleto{(l)\mathstrut}{6pt}}_{\mu\nu}\,,
\ee
where $\delta_{\mu\nu}$ is the leading term corresponding to a vanishing background field. Using this generic expansion, one can thus apply standard perturbation methods in order to calculate up to the n$^{\text{th}}$ variation of \eqref{OneLoopGour} with respect to the inverse effective metric $M^{\mu\nu}$
\be\label{Gexp}
\sum_{l=0}^n\frac{1}{l!}\,\delta^l\Gamma_{1}^{\mathrm{div}}\biggr\rvert_{\scaleto{M^{\mu\nu}=\delta^{\mu\nu}\mathstrut}{6pt}}\,.
\ee

The connection to a specific theory is then done by interpreting the expression of the effective inverse metric \eqref{expSOO} as well as a perturbative expansion in $ \bar{\pi}$
\be\label{Minvexp}
M^{\mu\nu}=\delta^{\mu\nu}+\sum_{l=1}^n \frac{1}{l!}H_{l}^{\mu\nu}\,.
\ee
For the theory at hand, the series stops at the second order and the explicit expressions are
\begin{IEEEeqnarray}{rCl}\label{Hdef}
H_1^{\mu\nu}&=&12\frac{c_3}{\Lambda^3}\,\left[\partial^\mu\partial^\nu \bar{\pi}-\delta^{\mu\nu}\,\Delta \bar{\pi}\right]\nonumber\\
H_2^{\mu\nu}&=&-72\frac{c_4}{\Lambda^2}\,\delta^{\mu\nu}\, \bar{\pi}^2 \\
H_{l>2}^{\mu\nu}&=&0\,.\nonumber
\end{IEEEeqnarray}
In that way, the series in \eqref{Gexp} can make contact with the specific theory at hand by relating the two expansions \eqref{Mexp} and \eqref{Minvexp} to each other at each order. For example, the first two relations are
\begin{IEEEeqnarray}{rCl}\label{Hhrel}
h^{\scaleto{(1)\mathstrut}{6pt}\,\mu\nu}&=&-H_1^{\mu\nu}\nonumber\\
h^{\scaleto{(2)\mathstrut}{6pt}\,\mu\nu}&=&2H_1^{\mu\rho}H_{1\,\rho}^{\nu}-H_2^{\mu\nu}\,.
\end{IEEEeqnarray}

This allows us to obtain the divergent one-loop contributions to the effective action of any order $\Gamma_{1,i}^{\mathrm{div}}$ by inserting \eqref{Hdef} into the expansion \eqref{Gexp} and extracting the term with the desired number $i$ of background fields $ \bar{\pi}$. Note that the potential term \eqref{pot} already contributes at the lowest order in the metric expansion \eqref{Gexp} with two background fields through the operator $P$ \eqref{expSOOP}. In doing so, we again recover exactly the same results \eqref{finalResults} of the previous section.


\section{Feynman diagrams}\label{sec_FeynDiag}

Additionally, we offer here a discussion of the individual one-loop Feynman diagrams, which will give direct access to the $\overline{\text{MS}}$-counterterms.
We will then make the link between the previous one-loop effective action computations and the explicit Feynman diagram results which closes the picture and serves as a complementary check of our calculations. 

We will therefore now calculate the divergent part of the one-loop $1$PI Feynman diagrams up to four external legs. 
Each diagram represents a contribution to the reduced matrix element $\mathcal{A}$ in the perturbative expansion of the S-matrix:
\be\label{Smatrix}
\bra{k_{\text{out}}}\mathcal{S}\ket{k_{\text{in}}}\bigg\rvert_{1\rm PI}=1+(2 \pi)^4\,\delta^4(k_{\text{out}}-k_{\text{in}})\, \mathcal{A}\;.
\ee
The reduced matrix element is calculated by summing over all possible Wick contractions of the form: 
\begin{equation}\label{Lagrangians}
\begin{split} 
& \contraction{}{\pi}{(x)}{\pi}
 \pi(x)  \bar{\pi}(y)=D(x-y)\\
& \bcontraction{}{\pi}{(x)}{ket{k}}
 \pi(x)\ket{k} =\mr{e}^{-ikx} \\
& \bcontraction{}{ket{k}}{}{\pi}
\bra{k} \pi(x) = \mr{e}^{ikx} \, ,
\end{split}
\end{equation}
where
\be
D(x-y)=\int \frac{\mr{d}^4p}{(2 \pi)^4}\,\mr{e}^{ip(x-y)}\; \, \frac{1}{p^2}
\ee
is the propagator of the massless scalar field with implicit Feynman-prescription.

Following the $\overline{\text{MS}}$-scheme, the one-loop counterterms can then be inferred from the UV divergence of the $1$PI diagrams which we will again extract using a dimensional regularization procedure with $d=4-2\epsilon$. We are thus after the log-divergent part of the one-loop $1$PI diagrams with $n$ external legs $\mathcal{A}_n^{\text{div}}$ which will be a function of the external momenta $k_i$, ${\scriptstyle i=1,..,n-1}$, since the overall delta-function $\delta^4(k_{\text{out}}-k_{\text{in}})$ always allows to express one momentum $k_n$ in terms of the others. For consistency we will treat all momenta as incoming throughout this section.
In the following, we will calculate all the divergent one-loop off-shell contributions up to four external legs.

\para{\bf 1-point function} 
Since, we have only the interaction in $\mathcal{L}_{3}$ with three legs, there is only one 1-point diagram at 1-loop order, namely the tadpole as shown in Fig. \ref{Feynman_diagrams_1} below.
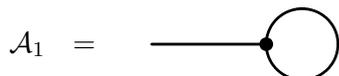
\begin{figure}[H]\vspace{20pt}
	\begin{center}
		$\mathcal{A}_1\ \ =$\hspace{10pt}
		\begin{fmffile}{Scattering2}
			\parbox{50mm}{\begin{fmfgraph*}(70,30)
					\fmfleft{i1}
					\fmfright{o1}
					\fmflabel{$ \bar{\pi}$}{i1}
					\fmf{plain}{i1,v1}
					\fmfdot{v1}
					\fmf{plain,left=1,tension=0.8,label=$ \bar{\pi}$}{v1,o1}
					\fmf{plain,left=1,tension=0.8}{o1,v1}
			\end{fmfgraph*}}
		\end{fmffile}
	\end{center}
	\caption{The tadpole contribution coming from $\mathcal{L}_{3}$.}
	\label{Feynman_diagrams_1}
\end{figure}
\noindent However, due to the antisymmetric structure of the interaction and the number of derivatives per field involved, the tadpole contribution vanishes identically.
\begin{align}
\mathcal{A}_{1}={}&0\,.
\label{eq:1ptmatter}
\end{align}

\begin{figure}[H]
\begin{center}
$\mathcal{A}_2\ \ =$\hspace{15pt}
\begin{fmffile}{Scattering2pf}
\parbox{20mm}{\subfloat[]{\begin{fmfgraph*}(50,50)
	           \fmfleft{i}
	            \fmfright{o} 
	            \fmf{plain}{i,v1} 
	            \fmf{plain}{v2,o}
                \fmfdot{v1,v2}
                \fmf{plain,left,tension=.3}{v1,v2,v1}
\end{fmfgraph*}}}+
\parbox{20mm}{\subfloat[]{\begin{fmfgraph*}(50,50)
	            \fmfsurround{i1,i2}
                \fmf{plain}{i1,v1}
                \fmf{plain}{i2,v1}
                \fmfdot{v1}
                \fmf{plain,right=0.5,tension=0.7}{v1,v1}
\end{fmfgraph*}}} \\[10pt]
\end{fmffile}
\end{center}
\caption{One-loop contributions to the 2-point function originating from (a) the cubic Galileon interaction $\mathcal{L}_{3}$ and (b) the Galileon breaking interaction $\mathcal{L}_{4}$.}
\label{Feynman_diagrams_2pt}
\end{figure}
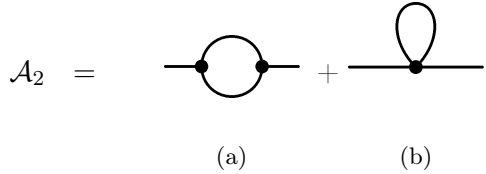

\para{\bf 2-point function} At 1-loop order, there are only the two diagrams shown in Fig. \ref{Feynman_diagrams_2pt} that contribute to the 2-point function.
The first diagram is the standard Galileon diagram coming from the $\mathcal{L}_{3}$ interactions
\begin{align}
\mathcal{A}_{2a}^{\rm div}=& -9\,\frac {\tilde{c}_3^2}{\Lambda^6} \int  \frac{\d^d p}{(2 \pi)^d} \frac{9 (p\cdot k_1)^4-16 p^2k_1^2(p\cdot k_1)^2+9p^4k_1^4}{p^2\; (p-k_1)^2}\biggr\rvert_{\rm div}   \nonumber\\
=&-\frac{1}{16 \pi^2\epsilon}\,\frac{\tilde{c}_3^2}{\Lambda^6}\;\frac{9}{4}p^8 \,,
\end{align}
On the other hand the tadpole type diagram \ref{Feynman_diagrams_2pt}(b) simply gives zero in dimensional regularization
\begin{align}
\mathcal{A}_{2b}=& \;36\,\frac {\tilde{c}_4}{\tilde{\Lambda}^2}\int  \frac{\d^d p}{(2 \pi)^d} \frac{p^2+k_1^2}{p^2} =0  \,.
\end{align}

\noindent Only starting from the 3-point function onward things will get interesting.

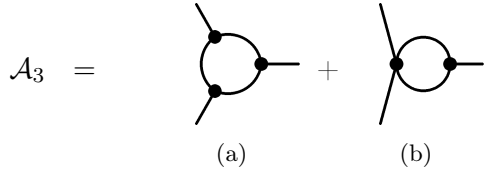
\begin{figure}[H]\vspace{10pt}
\begin{center}
$\mathcal{A}_3\ \ =$\hspace{15pt}
\begin{fmffile}{Scattering3pf}
\parbox{20mm}{\subfloat[]{\begin{fmfgraph*}(50,50)
	            \fmfsurround{i1,i2,i3}
                \fmf{plain}{i1,v1}
                \fmf{plain}{i2,v2}
                \fmf{plain}{i3,v3}
                \fmfdot{v1,v2,v3}
                \fmf{plain,right=0.5,tension=0.4}{v1,v2,v3,v1}
\end{fmfgraph*}}}+
\parbox{20mm}{\subfloat[]{\begin{fmfgraph*}(50,50)
	            \fmfsurround{i1,i2,i3}
                \fmf{plain}{i1,v1}
                \fmf{plain}{i2,v2}
                \fmf{plain}{i3,v2}
                \fmfdot{v1,v2}
                \fmf{plain,left,tension=0.3}{v1,v2,v1}
\end{fmfgraph*}}} \\[20pt]
\end{fmffile}
\end{center}
\caption{One-loop contributions to the 3-point function originating from (a) solely the cubic Galileon interaction and (b) a mixing between the Galileon and the Galileon breaking interaction.}
\label{Feynman_diagrams_3pt}
\end{figure}

\para{\bf  3-point function} There are again only two distinct $1$PI contributions with three external legs at one-loop order as shown in Fig. \ref{Feynman_diagrams_3pt}.

The first diagram \ref{Feynman_diagrams_3pt}(a) arises purely from the Galileon interaction and is therefore of no particular interest to us. The relevant contribution will come from the diagram \ref{Feynman_diagrams_3pt}(b). There are three distinct channels which need to be considered. For each of these channels at fixed vertices, there are apriori $72$ different ways of contracting in the S-matrix expansion \eqref{Smatrix} or in other words $3!4!$ different ways of distributing the $\mathcal{L}_{3}$ and $\mathcal{L}_{4}$ insertions over the legs in \ref{Feynman_diagrams_3pt}(b) divided by the symmetry factor of two. Note that for vertices with a different number of legs there is no additional vertex exchange factor which could cancel the $1/2!$ in the exponential expansion in \eqref{Smatrix}. The final result reads

\begin{IEEEeqnarray}{rCl}\label{Fresultc3c4}
\mathcal{A}_{3b}^{\rm div} =& -\frac{54}{16 \pi^2\epsilon}\,\frac{\tilde{c}_3\tilde{c}_4}{\Lambda^3\tilde{\Lambda}^2}\left[\right.&\left.2k_1^6+5 k_1^4k_2^2+6 k_1^4k_{12}+4 k_1^2k_{12}^2+8k_1^2k_2^2k_{12}+4 k_2^2k_{12}^2+6 k_{2}^4k_{12}
\right.\nonumber \\
&&\left. 
+5k_1^2 k_2^4+2k_2^6\right]\,
\end{IEEEeqnarray}
where we denote $k_{ij}\equiv k_i\cdot k_j$. Note that the result is symmetric under the exchange of momenta $k_1\leftrightarrow k_2$ as it should be.

\para{\bf  4-point function} At one-loop with four external legs, there are three distinct contributions but one is again a pure Galileon result coming solely from $\mathcal{L}_{3}$. The two interesting diagrams involving $\mathcal{L}_{4}$ are depicted in Fig. \ref{Feynman_diagrams_4pt}(b) and (c).

The first diagram \ref{Feynman_diagrams_4pt}(b) comes in a total of six different channels. Let's also quickly go through the combinatorics: For each of the six channels and fixed vertices there are $3!^24!=864$ different ways of distributing the insertions over the legs, since the symmetry factor of the diagram \ref{Feynman_diagrams_4pt}(b) is one. Vertex exchange of the two $\mathcal{L}_{3}$ insertions then introduces an additional factor of $2!$. Added up, the log-divergent part is calculated to be
\begin{IEEEeqnarray}{rCl}\label{Fresultc3c3c4}
\mathcal{A}_{4b}^{\rm div} =&-\frac{648}{16 \pi^2\epsilon}\,\frac{\tilde{c}_3^2\tilde{c}_4}{\Lambda^6\tilde{\Lambda}^2}\left[\right.
&\left.k_i^8-\tfrac{9}{2} k_i^6k_{j}^2+4 k_i^6k_{ij}+6 k_i^4k_{ij}^2+\tfrac{44}{3} k_i^4 k_{ij}k_{il}+\tfrac{16}{3} k_i^4 k_{ij}k_{jl}-\tfrac{2}{3} k_i^4 k_{jl}^2
\right. \nonumber \\
&&\left.
-\tfrac{27}{2} k_i^4k_j^2 k_{ij}-\tfrac{73}{6} k_i^4k_j^2 k_{il}-\tfrac{83}{6} k_i^4k_j^2 k_{jl}-\tfrac{26}{3} k_i^4k_j^4-19 k_i^4k_j^2k_l^2
\right. \nonumber \\
&&\left.
+4 k_i^2 k_{ij}^3+\tfrac{52}{3} k_i^2 k_{ij}^2k_{il}+\tfrac{13}{3} k_i^2 k_{ij}^2k_{jl}+\tfrac{29}{3} k_i^2 k_{ij}k_{jl}^2+\tfrac{56}{3} k_i^2 k_{ij}k_{il}k_{jl}
\right. \nonumber \\
&&\left.
-\tfrac{29}{3} k_i^2k_j^2 k_{ij}^2-\tfrac{38}{2} k_i^2k_j^2 k_{il}^2-\tfrac{52}{3} k_i^2k_j^2 k_{ij}k_{il}-\tfrac{70}{3} k_i^2k_j^2 k_{il}k_{il}-\tfrac{109}{3} k_i^2k_j^2k_l^2 k_{ij}
\right. \nonumber \\
&&\left.
+\tfrac{10}{3}  k_{ij}^4+ 14 k_{ij}^3k_{il}+\tfrac{64}{3} k_{ij}^2k_{il}^2+\tfrac{112}{3} k_{ij}^2k_{il}k_{jl}
\right.]
\end{IEEEeqnarray}
where this should be read as a sum over all $i,j,l=1,2,3$ but $i\neq j\neq l$. This of course reflects again the symmetry under exchange of momenta $k_1$, $k_2$ and $k_3$.

The contribution \ref{Feynman_diagrams_4pt}(c) with two insertions of $\mathcal{L}_{4}$ on the other hand comes in three different channels, each of which gets a contribution of $4!^2/2=288$ due to the symmetry factor of the diagram together with an additional vertex exchange contribution $2!$ which here just cancels the expansion coefficient $1/2!$ at second order. The total gives rise to a divergent part of the form 
\begin{IEEEeqnarray}{rCl}\label{Fresultc4c4}
\mathcal{A}_{4c}^{\rm div} =& -\frac{1296}{16 \pi^2\epsilon}\,\frac{\tilde{c}_4^2}{\tilde{\Lambda}^4}\left[\right.&\left.k_i^4+2 k_i^2k_{jl}+3 k_i^2k_j^2\right]\,.
\end{IEEEeqnarray}
where again $i,j,l=1,2,3$ but $i\neq j\neq l$.

\begin{figure}[H]\vspace{10pt}
\begin{center}
$\mathcal{A}_4\ \ =$\hspace{15pt}
\begin{fmffile}{Scattering4pftot}
\parbox{20mm}{\subfloat[]{\begin{fmfgraph*}(50,50)
	            \fmfleft{i1,i2}
	           \fmfright{o3,o4}
                \fmf{plain}{i1,v1}
                \fmf{plain}{i2,v2}
               \fmf{plain}{v3,o3}
                \fmf{plain}{v4,o4}
                \fmfdot{v1,v2,v3,v4}
                \fmf{plain,left=0.4,tension=0.7}{v1,v2,v4,v3,v1}
\end{fmfgraph*}}}+
\parbox{20mm}{\subfloat[]{\begin{fmfgraph*}(50,50)
	            \fmfleft{i1,i2}
	           \fmfright{o3,o4}
                \fmf{plain}{i1,v1}
                \fmf{plain}{i2,v2}
               \fmf{plain}{v3,o3}
                \fmf{plain}{v3,o4}
                \fmfdot{v1,v2,v3}
                \fmf{plain,left=0.5,tension=0.4}{v1,v2,v3,v1}
\end{fmfgraph*}}}+
\parbox{20mm}{\subfloat[]{\begin{fmfgraph*}(50,50)
	             \fmfleft{i1,i2}
	           \fmfright{o3,o4}
                \fmf{plain}{i1,v1}
                \fmf{plain}{i2,v1}
               \fmf{plain}{v2,o3}
                \fmf{plain}{v2,o4}
                \fmfdot{v1,v2}
                \fmf{plain,left,tension=0.3}{v1,v2,v1}
\end{fmfgraph*}}} \\[20pt]
\end{fmffile}
\end{center}
\caption{One-loop contributions to the 4-point function coming from (a) four insertions of $\mathcal{L}_{3}$, (b) the mixing between $\mathcal{L}_{3}$ and $\mathcal{L}_{4}$ and (b) from two insertions of $\mathcal{L}_{4}$.}
\label{Feynman_diagrams_4pt}
\end{figure}

The divergent part of the one-loop effective action of both the Schwinger-DeWitt and the geometrical computation can be compared to the off-shell Feynman diagram results by making use of the generating functional property of the effective action for one-loop $1$PI $n$-point correlation functions. Thus, Fourier-transformed functional derivatives of $ \Gamma_{1}^{\rm div}$ with respect to the background fields should coincide with the one-loop calculations of the diagramatic method. Details are spelled out in the appendix \ref{CrossCheck}.

We have explicitly checked that indeed all the effective action calculations \eqref{finalResults} match their corresponding off-shell $\overline{\text{MS}}$-counterterm calculated through Feynman diagram techniques. In particular, the results \eqref{resultc3c4} and \eqref{resultc4c4} perfectly agree with \eqref{Fresultc3c4} and \eqref{Fresultc4c4} respectively upon performing the transformations. The expression proportional to $\sim \tilde{c}_3^2\tilde{c}_4$ in \eqref{finalResults} matches also perfectly well the calculation \eqref{Fresultc3c3c4}.

The comparison and matching of the different methods provides a powerful check of our results, as they rely on fundamentally distinct concepts. Especially the agreement between the Feynman calculations and computations relying on the effective action can hardly be a coincidence, as the only common ground is the input of the Lagrangian.


\section{Conclusions}
\label{SecCon}
Combining theoretical constraints together with implications of cosmological observations is a powerful tool to discriminate EFTs or break degeneracies among them. 
In light of the GW170817 constraint, we considered dark energy models in the Horndeski framework with luminal speed of propagation for GWs and studied their radiative stability.
EFTs undergo a significant relative tuning of their classical operators in order to avoid ghostly propagating degrees of freedom, and in addition their overall coefficients 
have to satisfy given observational constraints. Therefore, their technical naturalness is an important theoretical prerequisite for their viability as an EFT. 
It is well known that the scalar Galileon model is technically natural and satisfies a non-trivial non-renormalization theorem. The Galileon operators are protected from quantum corrections and shift symmetric Horndeski models, so-called `weakly broken Galileons' \cite{Pirtskhalava:2015nla}, are also known to inherit some of these properties (quantum corrections are parametrically suppressed in these models). 

In the Horndeski survival model that we have considered here, the shift and the Galileon symmetry are broken due to an explicit dependence 
of the Horndeski $G_4$ function on the scalar field. Even though one naively would have expected to have lost any promising property of the radiative stability, we
were able to establish a well defined non-renormalization theorem also in this case. This is a very remarkable result. Using power counting arguments we
placed the
foundations of our non-renormalization theorem and introduced the involved classical and quantum expansion parameters. We then consolitated our power counting arguments
with the explicit calculation of quantum corrections using both a perturbative generalized Schwinger-DeWitt technique as well as a heat-kernel technique via geometrical resummation.
The latter served us as a generating functional for arbitrary n-point counterterms. These two techniques gave exactly matching results. 
Furthermore, we tested our
results for the one-loop effective action against the computation of individual Feynman diagrams and found full agreement. Our obtained one-loop counterterms coincide 
with our power-counting arguments. With this we have shown that the EFT of the specific Horndeski survival model is well defined on all relevant scales.

Going forward, in order to optimally test dark energy theories, it will be essential to better understand the interplay between theoretical priors associated with the viability of the underlying theory, as discussed here, and novel as well as established observational bounds. 
Exploring the impact of radiative stability related results (such as those established here) in conjunction with well-established observational bounds along the lines of \cite{Noller:2018eht} will be an obvious next step along this direction. 
Constraining radiatively stable theories with additional theoretical bounds is a further promising avenue.  
One example of such novel bounds are constraints from requiring the absence of dark energy instabilities induced by gravitational wave backgrounds (as ubiquitously sourced by inspiralling binary systems) \cite{Creminelli:2019kjy}. These already lead to significantly tightened parameter constraints on dark energy models when combined with bounds from cosmological observations (cf. constraints from \cite{Noller:2020afd} vs. those of \cite{Noller:2018wyv}) and it will be fascinating to further explore the resulting subset of theories. To this end note that Galilean symmtery breaking interactions of the type considered here are precisely those identified as a select few `survivors' by the constraints of \cite{Creminelli:2019kjy}, so our results identify a specific radiatively stable such survivor model.

More generally, as mentioned above, studying the theoretical viability of a specific EFT under consideration is an indispensable tool to constrain dark energy models.
One essential requirement is the radiative stability of the classical operators, which we focused on here. It is worth mentioning that the requirement of a unitary, causal and local UV completion can equally impose strong restrictions on dark energy models, especially when pairing the resulting positivity bounds with observational constraints (see \cite{Melville:2019wyy} and references therein). Other important theoretical restrictions for dark energy theories arise from Swampland conjectures, where the operators of a given theory have to satisfy some given upper bounds in order to be embeddable into a quantum gravity theory \cite{Agrawal:2018own,Heisenberg:2018yae}.

\acknowledgments
LH is supported by funding from the European Research Council (ERC) under the European Unions Horizon 2020 research and innovation programme grant agreement No 801781 and by the Swiss National Science Foundation grant 179740. JN is supported by an STFC Ernest Rutherford Fellowship, grant reference ST/S004572/1, and also acknowledges support from Dr. Max R\"ossler, the Walter Haefner Foundation and the ETH Zurich Foundation. 
\newpage


\appendix\label{appendix}

\section{Explicit geometrical objects}\label{explicityGeom}
Expanding the Levi-Civita symbols and canonically normalizing by setting $\tilde{c}_2=-\frac{1}{12}$, the effective inverse metric \eqref{expSOO} reads
\begin{align}\label{Minv}
M^{\mu\nu}={}&\delta^{\mu\nu} + 12 \frac{ \tilde{c}_3}{\Lambda^3}\,\left[\delta^{\mu \nu } \Delta \bar{\pi} + \partial^{\nu }\partial^{\mu }  \bar{\pi}\right]-  36\frac{\tilde{c}_4}{\tilde{\Lambda}^2}\,\delta^{\mu\nu}   \bar{\pi}^2 \,.
\end{align}
By treating this as a perturbation of the Euclidean metric in powers of the field $\bar{\pi}$ as $M^{\mu\nu}=\delta^{\mu\nu}+M_{\bar{\pi}}^{\mu\nu}$, this expression can be perturbatively inverted in a Neuman series schematically of the form $(I+M_{\bar{\pi}})^{-1}=\sum_{n=0}^\infty \,(-1)^n M^n_{\bar{\pi}}$ which yields
\begin{IEEEeqnarray}{rCl}\label{Meff}
M_{\mu\nu} &=&\delta_{\mu\nu} - 12\frac{\tilde{c}_3}{\Lambda^3}\left[ \delta_{ab} \Delta \bar{\pi} +  \partial_{b}\partial_{a}  \bar{\pi}\right] +36 \frac{ \tilde{c}_4}{\tilde{\Lambda}^2}\delta_{\mu\nu}   \bar{\pi}^2+144\frac{\tilde{c}_3^2}{\Lambda^6}\left[\delta_{\mu \nu } (\Delta\bar{\pi})^2 + \partial_{a}\partial_{\nu }  \bar{\pi} \partial_{\mu }\partial^{a}  \bar{\pi} \right.
\nonumber \\ 
&&  
\left.+ 2 \Delta \bar{\pi} \partial_{\nu }\partial_{\mu }  \bar{\pi}\right] -  864 \frac{ \tilde{c}_3 \tilde{c}_4 }{\Lambda^3\tilde{\Lambda}^2} \left[\delta_{\mu \nu } \Delta \bar{\pi}   \bar{\pi}^2 +   \bar{\pi}^2 \partial_{\nu}\partial_{\mu }  \bar{\pi}\right]-1728 \frac{ \tilde{c}_3^3 }{\Lambda^9}\left[\delta_{\mu \nu } (\Delta\bar{\pi})^3\right. 
\nonumber \\ 
&& 
\left. + 3 \Delta \bar{\pi} \partial_{a}\partial_{\nu }  \bar{\pi} \partial_{\mu }\partial^{a}  \bar{\pi} + \partial_{b}\partial_{a}  \bar{\pi} \partial_{\mu }\partial^{a}  \bar{\pi} \partial_{\nu }\partial^{b}  \bar{\pi} + 3 (\Delta\bar{\pi})^2 \partial_{\nu }\partial_{\mu }  \bar{\pi} \right] +1296 \frac{ \tilde{c}_4^2}{\tilde{\Lambda}^4}\delta_{\mu\nu}   \bar{\pi}^4
\nonumber \\ 
&& 
+ 15552\frac{ \tilde{c}_3^2 \tilde{c}_4}{\Lambda^6\tilde{\Lambda}^2} \left[ \delta_{\mu \nu } (\Delta\bar{\pi})^2   \bar{\pi}^2 +   \bar{\pi}^2 \partial_{a}\partial_{\nu }  \bar{\pi} \partial_{\mu }\partial^{a}  \bar{\pi} + 2 \Delta \bar{\pi}  \bar{\pi}^2 \partial_{\nu }\partial_{\mu } \bar{\pi} \right]
\nonumber \\ 
&& 
 +20736\frac{ \tilde{c}_3^4 }{\Lambda^{12}}\left[ \delta_{\mu \nu } (\Delta \bar{\pi})^4 + 6 (\Delta \bar{\pi})^2 \partial_{a}\partial_{\nu } \bar{\pi} \partial_{\mu }\partial^{a} \bar{\pi} + 4 \Delta  \bar{\pi} \partial_{b}\partial_{a} \bar{\pi} \partial_{\mu }\partial^{a} \bar{\pi} \partial_{\nu }\partial^{b} \bar{\pi}  \right.
\nonumber \\ 
&& \left.+ \partial_{c}\partial_{b} \bar{\pi} \partial^{c}\partial_{a} \bar{\pi} \partial_{\mu }\partial^{a} \bar{\pi} \partial_{\nu }\partial^{b} \bar{\pi} + 4 (\Delta \bar{\pi})^3 \partial_{\nu }\partial_{\mu } \bar{\pi}\right]
+ {\cal O}( \bar{\pi}^5)\,.
\end{IEEEeqnarray}
where we explicitly show the expansion up to $4^{\text{th}}$ order in the field such that \eqref{defeffM} is satisfied up to fourth order. The effective determinant is then defined as
\begin{align}
M \equiv \frac{1}{4!}\epsilon^{\mu_1 \mu_2 \mu_3 \mu_4}\epsilon^{\nu_1 \nu_2 \nu_3 \nu_4} M_{\mu_1 \nu_1}M_{\mu_2 \nu_2}M_{\mu_3 \nu_3}M_{\mu_4 \nu_4},
\end{align}
Explicitly we find
\begin{IEEEeqnarray}{rCl}\label{detM}
M&=& 1 - 36 \frac{\tilde{c}_3}{\Lambda^3}\Delta  \bar{\pi}+144 \frac{\tilde{c}_4}{\tilde{\Lambda}^2}   \bar{\pi}^2 + 72 \frac{ \tilde{c}_3^2}{\Lambda^6}\left[11 (\Delta \bar{\pi})^2 + \partial_{b}\partial_{a} \bar{\pi} \partial^{b}\partial^{a} \bar{\pi} \right] -  6480 \frac{\tilde{c}_3 \tilde{c}_4}{\Lambda^3\tilde{\Lambda}^2} \Delta  \bar{\pi}  \bar{\pi}^2
\nonumber \\ 
&& 
- 288\frac{\tilde{c}_3^3}{\Lambda^9}\left[47 (\Delta \bar{\pi})^3 + 15 \Delta  \bar{\pi} \partial_{b}\partial_{a} \bar{\pi} \partial^{b}\partial^{a} \bar{\pi} + 2 \partial^{b}\partial^{a} \bar{\pi} \partial_{c}\partial_{b} \bar{\pi} \partial^{c}\partial_{a} \bar{\pi}\right]+12960 \frac{ \tilde{c}_4^2 }{\tilde{\Lambda}^4} \bar{\pi}^4
\nonumber \\ 
&&
+ 15552\frac{ \tilde{c}_3^2 \tilde{c}_4}{\Lambda^6\tilde{\Lambda}^2} \left[11 (\Delta \bar{\pi})^2  \bar{\pi}^2 +  \bar{\pi}^2 \partial_{b}\partial_{a} \bar{\pi} \
\partial^{b}\partial^{a} \bar{\pi}\right] +2592\frac{ \tilde{c}_3^4 }{\Lambda^{12}}\left[75 (\Delta \bar{\pi})^4\right.
 \nonumber \\
 && 
 \left.+ 58 (\Delta \bar{\pi})^2 \partial_{b}\partial_{a} \bar{\pi} \partial^{b}\partial^{a} \bar{\pi} + 16 \Delta  \bar{\pi} \partial^{b}\partial^{a} \bar{\pi} \partial_{c}\partial_{b} \bar{\pi} \partial^{c}\partial_{a} \bar{\pi} + 2 \partial^{b}\partial^{a} \bar{\pi} \partial^{c}\partial_{a} \bar{\pi} \partial_{d}\partial_{c} \bar{\pi} \partial^{d}\partial_{b} \bar{\pi} \phantom{\frac{1}{2}} \right.
 \nonumber \\
 && 
\left.+ \partial_{b}\partial_{a} \bar{\pi} \partial^{b}\partial^{a} \bar{\pi} \partial_{d}\partial_{c} \bar{\pi} \partial^{d}\partial^{c} \bar{\pi} \right]+ {\cal O}( \bar{\pi}^5)\,.
\end{IEEEeqnarray}

The effective metric \eqref{Meff} and it's inverse \eqref{Minv} can then be used in order to determine the various geometrical objects defined in \S\ref{Geom}. For example, the associated Christoffel symbols up to third order are given by
\begin{IEEEeqnarray}{rCl}
 \Gamma^{\rho}_{\mu\nu}(M)&=&\frac{M^{\rho\sigma}}{2}\left(\partial_{\mu}M_{\sigma\nu}+\partial_{\nu}M_{\sigma\mu}-\partial_{\sigma}M_{\mu\nu}\right)
 \nonumber \\ 
 &=&\scaleto{ 
 -6 \frac{ \tilde{c}_3}{\Lambda^3} \left[\delta_{\nu }{}^{\rho } \partial_{\mu }\Delta \bar{\pi} + \delta_{\mu }{}^{\rho } \partial_{\nu }\Delta \bar{\pi} -  \delta_{\mu \nu } \partial^{\rho }\Delta \bar{\pi} + \partial^{\rho }\partial_{\nu }\partial_{\mu }\bar{\pi}\right]+36\frac{ \tilde{c}_4 }{\tilde{\Lambda}^2} \left[\delta_{\nu }{}^{\rho } \pi \partial_{\mu }\pi+ \delta_{\mu }{}^{\rho } \pi \partial_{\nu }\pi -  \delta_{\mu \nu } \pi \partial^{\rho }\pi\right]
\mathstrut}{21pt}\nonumber \\ 
&& \scaleto{ 
+72\frac{c_{3}{}^2}{\Lambda^6} \left[\delta_{\nu }{}^{\rho } \Delta \bar{\pi} \partial_{\mu }\Delta \bar{\pi} + \delta_{\mu }{}^{\rho } \Delta \bar{\pi} \partial_{\nu }\Delta \bar{\pi} -  \delta_{\mu \nu } \Delta \bar{\pi} \partial^{\rho }\Delta \bar{\pi} - 2 \partial_{\nu }\partial_{\mu }\pi \partial^{\rho }\Delta \bar{\pi}+ \delta_{\mu \nu } \partial^{a}\Delta \bar{\pi} \partial^{\rho }\partial_{a}\pi  \right.
\mathstrut}{21pt}\nonumber \\ 
&& \scaleto{  
\left. + \partial_{a}\partial_{\nu }\partial_{\mu }\pi \partial^{\rho }\partial^{a}\pi+ \partial_{\nu }\Delta \bar{\pi} \partial^{\rho }\partial_{\mu }\pi + \partial_{\mu }\Delta \bar{\pi} \partial^{\rho }\partial_{\nu }\pi + \Delta \bar{\pi} \partial^{\rho }\partial_{\nu }\partial_{\mu }\pi\right]-216 \frac{c_{3}{} c_{4}{} }{\Lambda^3\tilde{\Lambda}^2}\left[\delta_{\nu }{}^{\rho } \Delta \bar{\pi} \partial_{\mu }\Delta \bar{\pi}\right.
\mathstrut}{19pt}\nonumber \\ 
&& \scaleto{ 
+ \delta_{\mu }{}^{\rho } \Delta \bar{\pi} \partial_{\nu }\Delta \bar{\pi} -  \delta_{\mu \nu } \Delta \bar{\pi} \partial^{\rho }\Delta \bar{\pi} - 2 \partial_{\nu }\partial_{\mu }\pi \partial^{\rho }\Delta \bar{\pi}+ \delta_{\mu \nu } \partial^{a}\Delta \bar{\pi} \partial^{\rho }\partial_{a}\pi + \partial_{a}\partial_{\nu }\partial_{\mu }\pi \partial^{\rho }\partial^{a}\pi \phantom{\frac{1}{2}}
 \mathstrut}{21pt}\nonumber \\ 
&& \scaleto{  
\left. + \partial_{\nu }\Delta \bar{\pi} \partial^{\rho }\partial_{\mu }\pi + \partial_{\mu }\Delta \bar{\pi} \partial^{\rho }\partial_{\nu }\pi+ \Delta \bar{\pi} \partial^{\rho }\partial_{\nu }\partial_{\mu }\pi  \right] - \frac{c_{3}{}^3 }{\Lambda^9}\left[\delta_{\nu }{}^{\rho } \Delta \bar{\pi}^2 \partial_{\mu }\Delta \bar{\pi} + \delta_{\mu }{}^{\rho } \Delta \bar{\pi}^2 \partial_{\nu }\Delta \bar{\pi}\right.
\mathstrut}{21pt} \nonumber \\ 
&&\scaleto{ 
-  \delta_{\mu \nu } \Delta \bar{\pi}^2 \partial^{\rho }\Delta \bar{\pi}- 3 \partial_{a}\partial_{\nu }\pi \partial_{\mu }\partial^{a}\pi \partial^{\rho }\Delta \bar{\pi} - 4 \Delta \bar{\pi} \partial_{\nu }\partial_{\mu }\pi \partial^{\rho }\Delta \bar{\pi} + 2 \delta_{\mu \nu } \Delta \bar{\pi} \partial^{a}\Delta \bar{\pi} \partial^{\rho }\partial_{a}\pi \phantom{\frac{1}{2}}
 \mathstrut}{21pt}\nonumber \\ 
&& \scaleto{ 
+ \partial_{\mu }\partial^{a}\pi \partial_{\nu }\Delta \bar{\pi} \partial^{\rho }\partial_{a}\pi + \partial_{\mu }\Delta \bar{\pi} \partial_{\nu }\partial^{a}\pi \partial^{\rho }\partial_{a}\pi+ 2 \partial^{a}\Delta \bar{\pi} \partial_{\nu }\partial_{\mu }\pi \partial^{\rho }\partial_{a}\pi + 2 \Delta \bar{\pi} \partial_{a}\partial_{\nu }\partial_{\mu }\pi \partial^{\rho }\partial^{a}\pi \phantom{\frac{1}{2}}
  \mathstrut}{21pt}\nonumber \\ 
&& \scaleto{ 
-  \partial_{\mu }\partial^{a}\pi \partial_{\nu }\partial^{b}\pi \partial^{\rho }\partial_{b}\partial_{a}\pi + \partial_{b}\partial_{a}\partial_{\nu }\pi \partial_{\mu }\partial^{a}\pi \partial^{\rho }\partial^{b}\pi+ \partial_{b}\partial_{a}\partial_{\mu }\pi \partial_{\nu }\partial^{a}\pi \partial^{\rho }\partial^{b}\pi \phantom{\frac{1}{2}}
 \mathstrut}{21pt}\nonumber \\ 
&& \scaleto{
 \left.+ 2 \Delta \bar{\pi} \partial_{\nu }\Delta \bar{\pi} \partial^{\rho }\partial_{\mu }\pi+ 2 \Delta \bar{\pi} \partial_{\mu }\Delta \bar{\pi} \partial^{\rho }\partial_{\nu }\pi + \Delta \bar{\pi}^2 \partial^{\rho }\partial_{\nu }\partial_{\mu }\pi \right]
+ {\cal O}( \bar{\pi}^4)\,.\phantom{\frac{1}{2}}\mathstrut}{21pt}
\end{IEEEeqnarray}

In practice, however, a convenient way to proceed is to expand all the objects in \eqref{OneLoopGour} in terms of the effective metric $M_{\mu\nu}$, it's inverse $M^{\mu\nu}$ and the determinant $M$ and plug in the expressions \eqref{Meff}, \eqref{Minv} and \eqref{detM} respectively, in order to obtain the desired divergent one-loop effective action expressions in terms of the scalar background field. Alternatively, one can also expand \eqref{OneLoopGour} in a perturbation series of the inverse metric and match it to \eqref{Minv} without ever referring to the explicit expression of the effective metric as explained in the main text in \S\ref{Geom}.

\section{Cross-check details}\label{CrossCheck}

The effective action is a $1$PI generating functional in the sense that repeatedly applying functional derivatives with respect to the background field yields $1$PI correlation functions
\be
\frac{\delta^n\Gamma[\bar{\pi}]}{\delta\bar{\pi}(x_1)...\delta\bar{\pi}(x_n)}\biggr\rvert_{\bar{\pi}=\langle\pi\rangle} =\langle \pi(x_1)...\pi(x_n) \rangle_{1\rm PI}\,.
\ee
The $1$PI correlation functions in turn are given by the sum of all connected $1$PI diagrams with $n$ external points by the usual cancellation of exponentiated disconnected diagrams in the numerator and denominator. Thus, fourier transformed functional derivatives of our divergent one-loop effective action results \eqref{finalResults} at vanishing mean field should coincide with the corresponding divergent off-shell results of the $1$PI diagrams calculated in \S\ref{sec_FeynDiag}.

As an explicit example consider the 3-point effective action contribution \eqref{resultc3c4} proportional to $\tilde{c}_3\tilde{c}_4$
\be
\Gamma_{1,3}^{\rm div}(\scaleto{ \tilde{c}_3\tilde{c}_4\mathstrut}{8pt}) = - \frac{54}{16 \pi^2\epsilon}\,\frac{\tilde{c}_3\tilde{c}_4}{\Lambda^3\tilde{\Lambda}^2}\,\int \mathrm{d}^4x \,  \bar{\pi}\,\Delta  \bar{\pi}\,\Delta^2  \bar{\pi}
\ee 
Taking three functional derivatives of this expression gives
\small
\be
\frac{\delta^3\Gamma_{1,3}^{\rm div}(\scaleto{ \tilde{c}_3\tilde{c}_4\mathstrut}{8pt})}{\delta\bar{\pi}(x_1)\delta\bar{\pi}(x_2)\delta\bar{\pi}(x_3)} = \frac{54}{16 \pi^2\epsilon}\,\frac{\tilde{c}_3\tilde{c}_4}{\Lambda^3\tilde{\Lambda}^2}\,\int \mathrm{d}^4x \left[  \delta^4(\scaleto{x-x_1\mathstrut}{8pt})\,\left(\pd_{x}^2 \, \delta^4(\scaleto{x-x_2\mathstrut}{8pt})\right)\left(\pd_{x}^4 \delta^4(\scaleto{x-x_3\mathstrut}{8pt})\right)+ 5\, \text{perm.}\right]
\ee
\normalsize
Fourier transforming results in replacing each delta function by an exponential dependence on the corresponding incoming momentum $\int \mathrm{d}^4x_j \mathrm{e}^{ik_jx_j}\delta(x-x_j)=\mathrm{e}^{ik_jx}$ yielding the overall momentum conservation $\int \mathrm{d}^4x \mathrm{e}^{i(k_1+k_2+k_3)x}=(2\pi)^4\delta^4(k_1+k_2+k_3)$ which matches the factor in the definition of the reduced matrix element \eqref{Smatrix}. The various derivatives $\pd_x^2$ which applied on the exponential factors exactly reproduce the momentum structure \eqref{Fresultc3c4} upon expressing $k_3$ in terms of the other momenta, showing the equivalence of the results. The same can be done for each term of our results \eqref{finalResults}, all matching their corresponding Feynman diagram calculation.

\newpage
\bibliographystyle{JHEP}
\bibliography{references} 
\end{document}